\documentclass[letter]{aa}
\usepackage{graphicx}
\usepackage{txfonts}

\def\Lsun{L$_\odot$}
\def\Msun{M$_\odot$}

\def\Heii{He\,{\sc ii}}

\def\Civ{C\,{\sc iv}}

\def\Lya{Ly$\alpha$}
\def\Ha{H$\alpha$}

\def\kms{km\,s$^{-1}$}

\def\Kkmspc{K~km\,s$^{-1}$\,pc$^2$}
\def\lsim{\mathrel{\rlap{\lower 3pt \hbox{$\sim$}} \raise 2.0pt \hbox{$<$}}}
\def\gsim{\mathrel{\rlap{\lower 3pt \hbox{$\sim$}} \raise 2.0pt \hbox{$>$}}}

\begin{document}

\authorrunning{Decarli et al.}
\titlerunning{Dust and molecular gas in enormous Ly$\alpha$ nebulae}

\title{A search for dust and molecular gas in enormous Ly$\alpha$ nebulae at $z\approx 2$}
\author{
Roberto Decarli\inst{1}, Fabrizio Arrigoni-Battaia\inst{2}, Joseph F.\ Hennawi\inst{3,4}, Fabian Walter\inst{4}, Jason X.\ Prochaska\inst{5,6}, Sebastiano Cantalupo\inst{7,8}. 
}
\institute{
INAF -- Osservatorio di Astrofisica e Scienza dello Spazio di Bologna, via Gobetti 93/3, I-40129, Bologna, Italy. \email{ roberto.decarli@inaf.it} \and
Max-Planck-Institut f\"{u}r Astrophysik, Karl-Schwarzschild-Str 1, 85748 Garching bei M\"{u}nchen, Germany\and
Department of Physics, Broida Hall, University of California, Santa Barbara Santa Barbara, CA 93106-9530, USA\and
Max-Planck Institut f\"{u}r Astronomie, K\"{o}nigstuhl 17, D-69117, Heidelberg, Germany\and
UCO/Lick Observatory, University of California \u2013 Santa Cruz, Santa Cruz, CA 95064, USA\and
Kavli Institute for the Physics and Mathematics of the Universe (Kavli IPMU), 5-1-5 Kashiwanoha, Kashiwa, 277-8583, Japan\and
Department of Physics, ETH Zurich, Wolgang-Pauli-Strasse 27, 8093, Z\"{u}rich\and
Dipartimento di Fisica, Universit\`{a} degli Studi di Milano-Bicocca, Piazza della Scienza 3, I-20126 Milano, Italy
}

\date{August 2020}
\abstract{Enormous Ly$\alpha$ nebulae, extending over 300--500\,kpc around quasars, represent the pinnacle of galaxy and cluster formation. Here we present IRAM Plateau de Bure Interferometer observations of the enormous Ly$\alpha$ nebulae `Slug' ($z$=$2.282$) and `Jackpot' ($z$=$2.041$). Our data reveal bright, synchrotron emission associated with the two radio--loud AGN embedded in the targeted nebulae, as well as molecular gas, as traced via the CO(3-2) line, in three galaxies (two sources in the Slug, and one in the Jackpot). All of the CO emission is associated with galaxies detected in their rest-frame UV stellar emission. The total mass in molecular gas of these three galaxies [$\sim (3-5)\times10^{10}$\,\Msun] is comparable with the total ionized gas mass responsible for the diffuse nebular emission. Our observations place limits on the molecular gas emission in the nebulae: The molecular gas surface density is $\Sigma_{\rm H2}<12-25$\,\Msun{}\,pc$^{-2}$ for the Slug nebula and $\Sigma_{\rm H2}<34-68$\,\Msun{}\,pc$^{-2}$ for the Jackpot nebula. These are consistent with the expected molecular gas surface densities, as predicted via photoionization models of the rest-frame UV line emission in the nebulae, and via \Lya{} absorption in the Jackpot nebula. Compared to other radio--loud quasars at $z>1$, and high-redshift radio--loud galaxies, we do not find any strong trends relating the molecular gas reservoirs, the radio power, and the \Lya{} luminosities of these systems. The significant step in sensitivity required to achieve a detection of the molecular gas from the nebulae, if present, will require a substantial time investment with JVLA, NOEMA, or ALMA.
}

\maketitle

\section{Introduction}

Diffuse nebulae of hydrogen Lyman-$\alpha$ (\Lya), extending over 50-100 kpc, have been observed around star forming galaxies \citep[e.g.,][]{steidel00, matsuda04, dey05, yang09, wisotzki16, wisotzki18, li19, herenz20}, high-$z$ radio galaxies \citep[e.g.,][]{keel99, venemans02, miley06, marqueschaves19}, and quasars \citep[e.g.,][]{husband15, borisova16, fab16, fab19, ginolfi18, farina19, lusso19, drake20, travascio20} up to $z\sim6.5$. These large reservoirs of cool ($T\sim10^4$ K) gas extend well into the circum-galactic medium. They may link the pristine intergalactic medium accreting onto forming galaxies with the material outflowing due to winds, radio jets, or feedback from star formation or nuclear activity. Despite their central role in shaping the growth and evolution of galaxies, Ly$\alpha$ nebulae are still poorly understood. The bright Ly$\alpha$ emission has been imputed to photoionization from active galactic nuclei \citep{geach09}; to fluorence and reflection of the light from an embedded quasar \citep{cantalupo12,cantalupo14}; to integrated emission from star-forming satellites galaxies, or to material spread in the circum-galactic medium by galactic winds \citep[see, e.g.,][]{steidel11}; to shock heating of the gas via galactic superwinds \citep{taniguchi00}; to cooling from cold accretion \citep{haiman00,fumagalli12,trebitsch16}. Furthermore, Ly$\alpha$ photons are likely affected by complex radiative transfer effects, at least on small scales ($<$10\,kpc) where resonant scattering seems to be important \citep[e.g.,][]{ao20}. 

Four enormous Ly$\alpha$ nebulae (ELANe) stand out because of their exceptional spatial extent ($300$--$500$\,kpc) and total Ly$\alpha$ luminosity $L_{\rm Ly\alpha}=(0.2-1.5)\times10^{45}$ erg\,s$^{-1}$: The `Slug' nebula, discovered around the quasar UM287 at $z=2.279$  \citep{cantalupo14}; The `Jackpot' nebula, embedding the $z=2.041$ quasar SDSS\,J0841+3921 \citep{hennawi15}; the `MAMMOTH-I' nebula in the BOSS1441 overdensity at $z=2.319$ \citep{cai17}; and the giant nebula around the quasar SDSS~J102009.99+104002.7 at $z$=3.164 \citep{fab18}. All of these enormous nebulae harbour several spectroscopically--confirmed galaxies, including a number of active galactic nuclei (AGN) and several Ly$\alpha$ emitting galaxies, thus marking some of the most prominent galactic overdensities at these redshifts. The extent of these nebulae exceeds the expected diameter of the dark matter halos of the most massive galaxies ($M_{\rm halo}\sim 10^{12.5}$ \Msun{}) at these redshifts. These immense systems thus represent unique laboratories to study galaxy and structure formation at the very high--mass end of the mass spectrum.

Observations combining Ly$\alpha$, \Civ{}, \Heii{}, and \Ha{} provide clues on the physical properties of the clouds responsible for the diffuse Ly$\alpha$ emission in these ELANe. Interpreting these diagnostics with the support of photoionization models \citep[see, e.g.,][]{hennawi13} points to the presence of a population of small ($R<20$ pc), sparse ($N_{\rm H}<10^{20}$\,cm$^{-2}$) but dense ($n_{\rm H}\gsim 3$\,cm$^{-3}$) clouds \citep{fab15,cai17,leibler18,cantalupo19}. The relatively low Ly$\alpha$/H$\alpha$ flux ratio $F_{\rm Ly\alpha}/F_{\rm H\alpha}=5.5\pm1.1$ currently observed in a source embedded in the Slug nebula appears broadly consistent with expectations for Case B recombination, thus disfavoring photon-pumping or scatter from the quasar broad line region as the main powering mechanism in this part of the nebula \citep{leibler18}.

A critical open question on the nature of ELANe is whether these clouds reach a gas density sufficient to form significant molecular gas. Indeed, dedicated cloud-crushing simulations \citep[e.g.,][]{mccourt18,gronke18,gronke20} and high resolution cosmological simulations \citep[e.g.,][]{hummels19} point towards a scenario in which such dense cold clouds can survive in a hot-halo environment even after shattering in a mist.

Out of the four ELANe known to date, only the MAMMOTH-I nebula has been observed in its molecular gas content, as traced by the carbon monoxide ($^{12}$C$^{16}$O, hereafter CO) ground rotational transition CO(1-0) \citep{emonts19}. In this source, large reservoirs of molecular gas appear associated with galaxies or group of galaxies in the core of the nebula \citep[see also][]{yang14}. Noticeably, the CO emission appears brighter in regions that are devoid of strong \Lya{} emission. This is in contrast with earlier results on the `Spiderweb' galaxy system, which shows atomic carbon, water vapor and CO emission widespread in the nebula \citep{emonts16,gullberg16}. \citet{emonts19} suggest that the discrepancy might be attributed to the presence of a strong, AGN-driven radio jet in the Spiderweb, that might cause local thermal instability that leads to gas cooling. In this scenario, the lack of a radio--loud source in the MAMMOTH-I system might explain the smaller reservoir of molecular gas with respect to the Spiderweb system.

\begin{figure*}
\begin{center}
\includegraphics[width=\columnwidth]{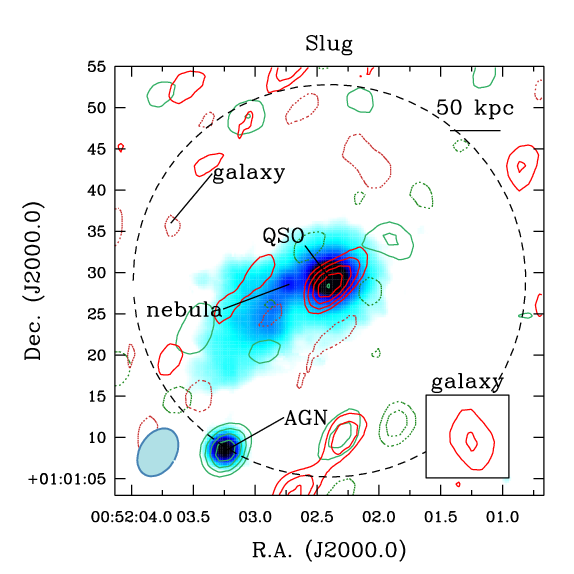}
\includegraphics[width=\columnwidth]{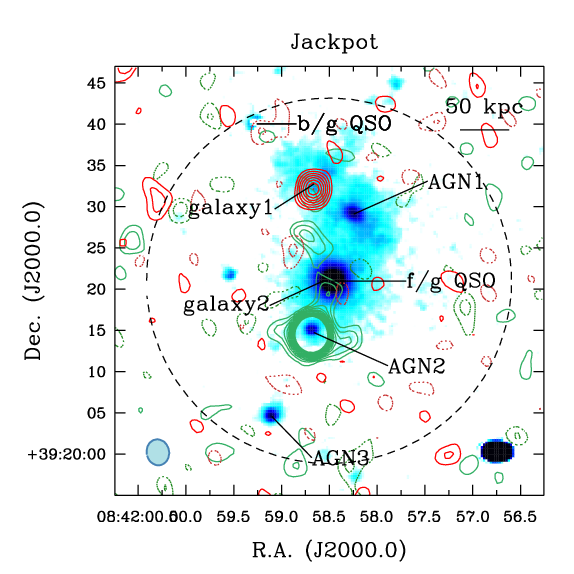}
\end{center}
\caption{The enormous Ly$\alpha$ nebulae `Slug' and `Jackpot', as seen in Ly$\alpha$ emission \citep[color scale; see][]{cantalupo14,hennawi15}, in CO(3-2) (red contours) and 3\,mm continuum (green contours). The large dashed circles show the primary beams of our 3\,mm observations, while the blue ellipses at the bottom left corner show the synthesized beams. The solid/dotted contours show the positive/negative 2,3,4,5,\ldots 15-$\sigma$ isophotes, where 1-$\sigma$=122 and 100$\,\mu$Jy\,beam$^{-1}$ for the CO(3-2) line maps of the Slug and Jackpot nebulae, respectively, and 1-$\sigma$=28 and 38$\,\mu$Jy\,beam$^{-1}$ for the 3\,mm continuum of the two systems. Line maps are integrated over 300\,\kms{} (see Fig.~\ref{fig_spec}). For the Slug, the inset shows the CO map of the galaxy on the eastern side. The synchrotron continuum of the two radio-loud AGN in the two nebulae are also secured at high significance. Conversely, the UV-luminous quasar at the center of the Slug and a star-forming galaxy at the northern edge of the Jackpot nebula are the only unambiguous CO detections in our data. No CO line nor 3mm continuum detection is reported in either nebula.}\label{fig_ima}
\end{figure*}

In order to investigate the molecular gas content of ELANe, we obtained 3\,mm interferometric observations of the Slug and Jackpot nebulae. We targeted the CO(3-2) transition and its underlying dust continuum with the IRAM Plateau de Bure Interferometer (PdBI) / Northern Extended Millimeter Array (NOEMA). Both these ELANe host a radio--loud AGN, a UV--luminous quasar, and other galaxies. This Letter is structured as follows: \S\ref{sec_obs} describes our CO(3-2) observations. In \S\ref{sec_results} we present our results. We discuss the implications of our findings in \S\ref{sec_conclusions}. 

Through this report we assume a $\Lambda$CDM cosmology, with $H_0=70$ \kms{}\,Mpc$^{-1}$, $\Omega_{\rm m}=0.3$, and $\Omega_{\Lambda}=0.7$, and a Kroupa initial mass function to compute star formation rates (SFR). In this cosmological framework, the scale distance at $z$=$2.041$ ($z$=$2.279$) is 8.35 kpc\,arcsec$^{-1}$ (8.22 kpc\,arcsec$^{-1}$), and the luminosity distance is 15.928\,Gpc (18.223\,Gpc).

\section{Observations}\label{sec_obs}

We searched for the CO(3-2) line in the Slug and Jackpot nebulae using PdBI. The line has a rest-frame frequency $\nu_{\rm 0}=345.796$ GHz. At $z=2.0-2.3$, it is redshifted into the 3\,mm transparent window of the atmosphere.

Observations of the Slug nebula were carried out in a number of short tracks during 2014 (May 13, 15, 18; June 01; October 17, 18, 19, 20, 21, 25, 26; program IDs: X0B1 and S14CH). Baselines ranged between 15 and 100\,m with 5 or 6 antennas (compact 5Dq and 6Dq configurations). The tuning frequency of our observations was 105.436\,GHz (WideX band 1). The pointing center was set to RA=00:52:02.400, Dec=+01:01:29.00 (J2000.0), i.e., centered on the bright quasar in the nebula. The primary beam of PdBI has a Gaussian profile with full width at half maximum (FWHM) = $47.8''$, at the tuning frequency of our observations, i.e., it is large enough to easily accommodate the nebula, and it reaches out to the companion radio-loud AGN (see Fig.~\ref{fig_ima}). The quasar 3C454.3 and the source MWC349 were used for bandpass and flux calibration, while the quasars 0106+013 and 0112-017 were repeatedly observed for the amplitude and phase calibrations. The precipitable water vapor was typically $\sim 5$\,mm, and with a minimum of $1-2$\,mm (May 15) and a maximum of 5--10\,mm (May 13). We processed our data using the June 2014 and October 2014 versions of the \textsf{GILDAS} software. The final cube consists of 16500 visibilities, corresponding to $13.75$\,hr on source (6-antennas equivalent). 

Observations of the Jackpot nebula started in 2014 (April 13, 14, 15; program ID: X0B1), and continued in 2017 (May 05, 08, 15, 19, 22; program ID: W16DA). The 2014 data consists of three short visits in 6C configuration. The 2017 data comprizes five observations in 8D configuration. Baselines ranged between 20--180\,m. The tuning frequency was set to 113.711\,GHz. The pointing center was set to RA=08:41:58.500, Dec=+39:21:21.00 (J2000.0). The primary beam at the tuning frequency is $44.3''$, thus fully encompassing the entirety of the nebula (see Fig.~\ref{fig_ima}). The quasars 3C84, 0851+202, and 0716+714 served as bandpass and flux calibrators, while the quasars 0821+394 and 0923+392 were observed as phase and amplitude calibrators. The precipitable water vapor during the observations ranged between 3 and 15\,mm. We processed our data using the June 2014 and May 2017 versions of the \textsf{GILDAS} software. The final data cube comprizes of 26428 visibilities, corresponding to $11.80$\,hr (8-antennas equivalent). 

We imaged the cubes using the \textsf{GILDAS} suite \textsf{mapping}. Natural weighting was adopted. The beam size is $6.24''\times4.38''$ ($\sim 51 \times 36$\,kpc$^2$ at $z=2.279$) for the Slug nebula, and $3.17''\times 2.76''$ ($\sim 26\times23$\,kpc$^2$) for the Jackpot. After the spectral resampling, we masked the channel centered at 103.928\,GHz in the Slug observations, due to a parasite frequency. We resampled the spectral dimension in 50\,\kms{} channels. We also created continuum maps by averaging over the entire observed spectral range, using the task \textsf{uv\_average} within \textsf{mapping}. We estimate an rms of $390$ $\mu$Jy\,beam$^{-1}$ per 50 \kms{} channel and $437$ $\mu$Jy\,beam$^{-1}$ per 50 \kms{} channel for the Slug and Jackpot observations, respectively. In the continuum and line maps, we applied the task \textsf{clean} down to 1.5-$\sigma$ in a cleaning box of a few pixels centered on the radio-loud AGN.

\begin{figure}
\begin{center}
\includegraphics[width=\columnwidth]{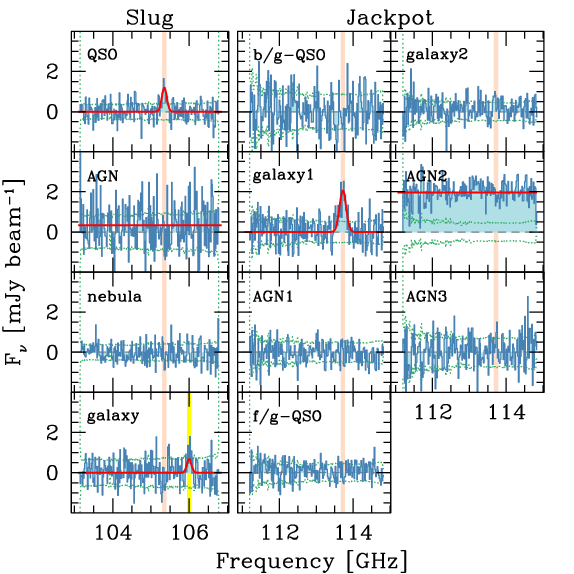}
\end{center}
\caption{Primary-beam corrected PdBI/NOEMA 3\,mm spectra of various sources embedded in the Slug and Jackpot nebulae (see their location in Fig.~\ref{fig_ima}). The adopted channel width is 50\,\kms{}. The dotted green histograms mark the 1-$\sigma$ range. 
Pink / yellow shades highlight the channels used in the line maps in Fig.~\ref{fig_ima}. The thick solid red lines show the fits for the three reported line emitters and two continuum emitters, as described in the text. All of the line detections are clearly associated with optical galactic counterparts.}\label{fig_spec}
\end{figure}

\section{Results}\label{sec_results}

Fig.~\ref{fig_ima} compares the CO and radio continuum maps of the targets of this study, with the continuum--subtracted Ly$\alpha$ maps of the ELANe from \citet{cantalupo14} and \citet{hennawi15}. Fig.~\ref{fig_spec} shows the 3\,mm spectra extracted at the position of various sources in the Slug and Jackpot ELANe, as marked in Fig.~\ref{fig_ima}.

\subsection{Continuum at 3\,mm}

We clearly detect the continuum emission of the radio--loud AGN (RL-AGN) in both the Slug (S/N$\approx$7) and the Jackpot (S/N$\sim$80) nebulae. No other source is detected in its continuum emission.

We fit the 3\,mm spectra of these sources with a constant flux density, using our custom Markov-Chain Monte Carlo routine \textsf{smc}. As prior, we adopt a wide Gaussian probability distributions centered on the continuum median value. Tab.~\ref{tab_fits} reports the fitted values.

If the 3\,mm continuum emission arises from dust thermal emission, the resulting integrated infrared luminosities, $L_{\rm IR}$, computed assuming a modified black body, is $L_{\rm IR}\gsim10^{13}$\,\Lsun{} for the AGN in the Slug ELAN, and $L_{\rm IR}\gsim10^{14}$\,\Lsun{} for the AGN in the Jackpot ELAN for any $T_{\rm dust}>35$\,K. These estimates place these sources among the most luminous IR galaxies. As dust acts as a catalyst for molecular gas formation, such high dust luminosities should be paired with exceptionally luminous CO emission ($>2$\,mJy, conservatively assuming a low, Milky Way--like CO excitation and a line width of 300\,\kms{}). Such a high CO luminosity is ruled out by our observations. We therefore conclude that the observed 3\,mm continuum emission in these two sources is of non-thermal origin. By comparing the 3\,mm flux density from our observations with the 1.4\,GHz flux density from FIRST \citep{white94}, as well as with observations at 4.9\,GHz and 8.4\,GHz by \citet{dipompeo11} for the Jackpot source, we infer radio spectral indexes $\alpha=-1.00 \pm 0.03$ for the radio-loud AGN in the Slug ELAN, and $\alpha=-0.461\pm0.009$ for the one in the Jackpot ELAN. No previous estimates of the radio spectral index in the Slug RL-AGN are available in the literature. On the other hand, our estimate of $\alpha$ for the RL-AGN in the Jackpot is in agreement with the $\alpha_{\rm fit}=-0.5$ value quoted in \citet{dipompeo11}.

\subsection{CO lines}

In the Slug nebula, we clearly detect CO(3-2) emission at S/N$\approx$7 associated with the UV--luminous quasar UM287. A tentative detection (S/N$\approx$4) is also reported at the position of a UV--detected galaxy at the Eastern edge of the sytem. In the Jackpot system, only a galaxy in the Northern part of the system is detected in CO(3-2) (S/N$\approx$13). 

We fit the lines with a Gaussian function, using \textsf{smc}. As priors, we assume loose Gaussian probability distributions centered on the ELAN redshift and on the maximum flux density of the line, as well as a Maxwellian probability distribution with scale width of 300\,\kms{} for the line widths. The results are listed in Tab.~\ref{tab_fits} and shown as red lines in Fig.~\ref{fig_spec}. 

The CO--based redshift of UM287 ($z$=$2.2824\pm0.0003$) is in excellent agreement with previous estimates based on rest-frame optical/UV lines \citep{leibler18,cantalupo19}, and slightly higher than the one reported by \citet{hewett10} ($z$=$2.279$). The line-of-sight velocity difference between the quasar and the other CO--emitting galaxy in the Slug is 1850\,\kms{}, hinting at a physical association of the galaxy with the overdensity hosting the ELAN, although we note that the velocity difference is larger (by a factor $\sim 2$) than the velocity gradient observed in the nebula based on \Ha{} emission \citep{leibler18}. The CO--detected galaxy in the Jackpot nebula has a redshift of $z$=$2.0407$. For comparison, the UV--bright `foreground' quasar in the system has a redshift of $z$=$2.04537\pm 0.00062$ \citep{hewett10}, i.e., at a line-of-sight velocity difference $\Delta v$=$460\pm 25$\,\kms{}, consistent with a physical association with the same overdensity.

We convert the line fluxes into line luminosities, $L'$, following, e.g., \citet{carilli13}:
\begin{equation}\label{eq_lum}
\frac{L'}{\rm K\,km\,s^{-1}\,pc^2} = \frac{3.25\times 10^7}{1+z} \frac{F_{\rm line}}{\rm Jy\,km\,s^{-1}}\,\left(\frac{\nu_{0}}{\rm GHz}\right)^{-2}\,\left(\frac{D_{\rm L}}{\rm Mpc}\right)^2
\end{equation}
where $F_{\rm line}$ is the line integrated flux, and $D_{\rm L}$ is the luminosity distance. We then estimate associated molecular gas masses as: $M_{\rm H2}$=$\alpha_{\rm CO}\,r_{31}^{-1}\,L'$, where we adopt a CO--to--H$_2$ conversion factor $\alpha_{\rm CO}$=3.6\,\Msun{}\,(\Kkmspc)$^{-1}$ from \citet{daddi10} \citep[see also][]{bolatto13}, and $r_{31}$=0.80 from the $z>2$ sample of galaxies in \citet{boogaard20}. Under these assumptions, we infer molecular gas masses of $(24\pm3)\times10^9$\,\Msun, $(13\pm3)\times10^9$\,\Msun{}, and $(48\pm4)\times10^9$\,\Msun{} for the quasar in the Slug, the galaxy on its eastern side, and for the CO--detected galaxy at the northern end of the Jackpot ELAN, respectively. Table \ref{tab_fits} lists all of the measured and inferred quantitities for these galaxies.

The line luminosities and associated molecular gas masses in our study are consistent with values reported in other quasar host galaxies at high redshift (see Fig.~\ref{fig_lya_radio_co} and the compilations in, e.g., \citealt{carilli13} and \citealt{venemans17}), as well as in other star forming galaxies at these redshifts \citep[e.g.,][]{tacconi18,aravena19}.

\begin{table*}
\caption{\rm Results from the 3\,mm spectral fitting of sources in the targeted ELANe.}\label{tab_fits}
\begin{center}
\begin{tabular}{c|ccc|cc}
\hline
                                       & \multicolumn{3}{c}{\em Slug}                                                      & \multicolumn{2}{c}{\em Jackpot}                      \\
Source                                 & QSO	     		    & AGN	              & galaxy			   & galaxy1			& AGN2  		  \\
\hline
R.A. (J2000)                           & 00:52:02.40		    & 00:52:03.24             & 00:52:03.68		   & 08:41:58.66		&  08:41:58.66  	  \\
Dec. (J2000)                           & +01:01:29.3		    & +01:01:08.4             & +01:01:36.0		   & +39:21:33.1		&  +39:21:14.7  	  \\
$z_{\rm CO}$                           &$2.2824_{-0.0003}^{+0.0003}$& ---                     &$2.2623_{-0.0006}^{+0.0006}$&$2.0407_{-0.0002}^{+0.0003}$& ---			  \\
FWHM$_{\rm CO}$         [\kms]         & $185_{-25}^{+22}$          & ---                     & $170_{-38}^{+40}$  	   & $237_{-18}^{+20}$  	& ---			  \\
$F_{\rm CO}$           [Jy\,\kms{}]    & $0.194\pm0.023$            & ---                     & $0.101\pm0.026$		   & $0.465\pm0.036$		& ---			  \\
$F_{\nu}$(3\,mm)         [mJy]	       & ---                        & $0.33_{-0.05}^{+0.04}$  &  ---                       & ---			&$1.967_{-0.025}^{+0.025}$\\
$L'_{\rm CO(3-2)}$  [\Kkmspc]      & $(5.4\pm0.6)\times10^9$    & ---                     & $(2.8\pm0.7)\times 10^9$   & $(10.6\pm0.8)\times 10^9$	& ---			  \\
$M_{\rm H2}$ [\Msun]               &  $(24\pm3)\times10^9$      & ---                     & $(13\pm3)\times 10^9$      & $(48\pm4)\times 10^9$	& ---			  \\
\hline
\end{tabular}
\end{center}
\end{table*}

We search for line emission beyond the pre--selected coordinates from known galaxies using the code \textsf{findclumps} \citep[see][]{decarli19}. Within the primary beam radii of the two ELAN pointings, we find only one additional line candidate with S/N$>$5 in the Slug ELAN: it is centered at R.A.=00:52:03.05 and Dec=+01:01:23.0 (J2000.0) and has a S/N=$5.15$. However, the line is statistically consistent with noise features, and it resides at the edge of the bandwidth ($\nu_{\rm obs}$=$106.55$\,GHz), thus suggesting that the candidate is not a real astrophysical source. No additional lines are found in the Jackpot ELAN at S/N$>$5.

\section{Discussion and conclusions}\label{sec_conclusions}

We presented 3\,mm observations of two enormous \Lya{} nebulae at $z\sim2$, the Slug and the Jackpot systems. Both nebulae harbour several galaxies, including a radio--loud and various radio--quiet AGN. We detect 3\,mm continuum emission in the two radio--loud AGN, and argue that it is dominated by the non-thermal synchrotron emission. We also detect CO line emission in three sources: the UV--bright quasar in the core of the Slug nebula, and two star-forming galaxies in the outskirt of the two ELANe. Using standard assumptions, we convert the CO line luminosities into molecular gas masses. 

\begin{figure}
\begin{center}
\includegraphics[width=\columnwidth]{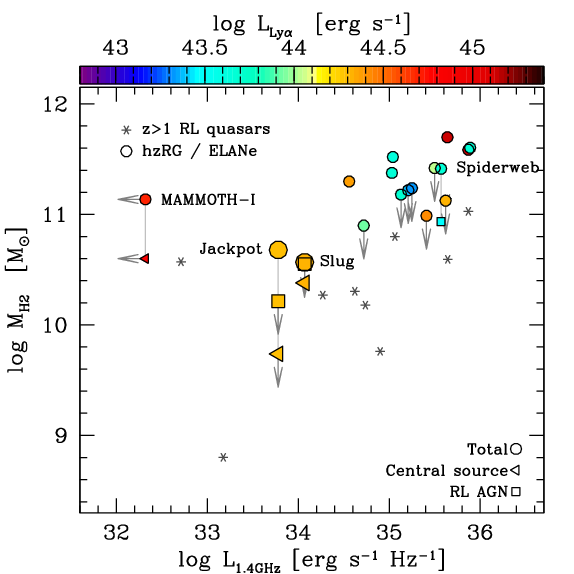}
\end{center}
\caption{Comparison of molecular gas mass (as traced via CO line emission), radio luminosity, and \Lya{} luminosity in ELANe, in high-redshift radio galaxies, and in $z>1$ radio loud quasars (see Sec.~\ref{sec_conclusions} for references). For the Spiderweb, MAMMOTH-I, Slug, and Jackpot systems, we show (i) the total molecular gas mass of the system (circles), (ii) the molecular gas mass of the radio-loud AGN, if any (squares), and (iii) the molecular gas mass of the source considered to be the main powering source of the ELAN (triangles). The apparent correlation between molecular gas and radio luminosity appears to reflect selection effects. No clear correlation is found between \Lya{} luminosity and other quantities. The lack of clear scaling relations highlight the complexity and diversity of these systems and of the occuring physical and radiative processes.}\label{fig_lya_radio_co}
\end{figure}

We find that the total mass in molecular gas in the two ELANe is $M_{\rm H2}$=$(37\pm4)\times10^9$\,\Msun{} in the Slug and $M_{\rm H2}$=$(48\pm4)\times10^9$\,\Msun{} in the Jackpot. These masses are comparable with the mass of cool ($T\sim 10^4$\,K) gas responsible for the nebular \Lya{} emission, albeit the former is concentrated in three individual galaxies, whereas the latter is spread on scales of several hundreds kpc. In Fig.~\ref{fig_lya_radio_co} we compare the molecular gas masses with the diffuse \Lya{} luminosity, and the radio luminosities of ELANe, of various high-redshift radio galaxies \citep{debreuck00,reuland03,miley08,emonts14,emonts19}, and of radio--loud quasars at $z>1$ \citep[from the compilation in][]{carilli13}. We do not find evidence of a correlation between \Lya{} luminosity and either CO or radio luminosity within the sampled ranges. We find an apparent correlation between molecular gas mass and radio luminosity; however this is likely a by-product of selection effects. Because radio selection historically drew the identification of these systems, which were then searched for CO emission, the faint--radio, bright--CO corner of the plot is undersampled, with the MAMMOTH-I being a clear example deviating from the apparent relation. At the same time, the detection of faint CO lines might be hindered in the presence of exceptionally bright synchrotron emission. We conclude that we find no evidence for a connection between radio luminosity and molecular gas content.

Under the same assumptions adopted for our line detections, and postulating a line width of 300\,\kms{}, the non-detection of CO associated with the nebulae implies a molecular gas mass limit between (1.8--3.6)$\times$10$^{10}$\,\Msun{}\,beam$^{-1}$ for the Slug nebula, (1.6--3.3)$\times$10$^{10}$\,\Msun{}\,beam$^{-1}$ for the Jackpot nebula, depending on the position within the primary beam (assuming 3-$\sigma$ significance). This translates into limits on the (beam--averaged, assuming beam areas of 1440\,kpc$^2$ and 470\,kpc$^2$ for the Slug and the Jackpot nebulae, respectively) surface molecular gas mass distribution $\Sigma_{\rm H2}<12-25$\,\Msun{}\,pc$^{-2}$ for the Slug nebula and $\Sigma_{\rm H2}<34-68$\,\Msun{}\,pc$^{-2}$ for the Jackpot nebula, or molecular gas column densities of $N_{\rm H2}<(0.75-1.6)\times 10^{21}$\,cm$^{-2}$ for the Slug and $N_{\rm H2}<(2.1-4.2)\times 10^{21}$\,cm$^{-2}$ for the Jackpot. Our limits exclude gas surface density regimes typical of starbursting environments. For comparison, the beam--averaged surface density of molecular gas measured in the circumgalactic medium of the Spiderweb by \citet{emonts16} is $\Sigma_{\rm H2}=35\pm11$\,\Msun{}\,pc$^{-2}$, i.e., comparable with the limits set by our observations in the Slug and Jackpot. The photoionization analysis presented by \citet{fab15} suggests a column density of $N_{\rm H}\lsim10^{20}$\,cm$^{-2}$, and \citet{hennawi15} estimate a column density of $N_{\rm H}\sim10^{20.4\pm0.4}$\,cm$^{-2}$ based on the \Lya{} absorption on the line--of--sight of the background quasar.

\citet{fab15} found that the nebulae likely host a population of dense ($n_{\rm H}\gsim 3$\,cm$^{-3}$), sparse clouds. The lack of 3mm continuum signal from the nebulae also place loose limits on the star formation occurring in these clouds. Assuming that the dust emission can be modeled as a modified black body with $T_{\rm dust}$=35\,K and $\beta$=1.6, we infer limits on the surface density of star formation of $\Sigma_{\rm SFR}<0.24$ and $0.79$ \Msun{}\,yr$^{-1}$\,kpc$^{-2}$, for the Slug and Jackpot nebulae, respectively. These limits exclude only rather extreme, starburst--like SFR surface densities; such intense star formation activity in the nebulae is already ruled out by the much tighter constraints that our observations place on $\Sigma_{\rm H2}$, provided that gas in the ELANe follows the `star formation law' observed in local galaxies \citep{bigiel08,leroy11,schruba11}.

Significantly deeper observations of molecular gas in ELANe are required in order to expose the molecular content (if present) of the diffuse material associated with the \Lya{} emission, and to conclusively distinguish between the different physical mechanisms that have been proposed to regulate the interaction between the radio jets and the growth and suppression of molecular gas reservoirs in ELANe and high--redshift radio galaxies. Such observations, despite expensive in terms of telescope time, may still be within reach with IRAM/NOEMA, ALMA, and JVLA.

\section*{Acknowledgments}
We thank the referee, B.~Emonts, for his feedback on the manuscript that allowed us to improve its quality. This work is based on observations carried out under projects X0B1, S14CH, W16DA with the IRAM Plateau de Bure / NOEMA Interferometer. IRAM is supported by INSU/CNRS (France), MPG (Germany) and IGN (Spain). The research leading to these results has received funding from the European Union's Horizon 2020 research and innovation program under grant agreement No 730562 [RadioNet].
We thank the IRAM staff and local contacts for their help and support in the data processing.
SC gratefully acknowledges support from the Swiss National Science Foundation grant PP00P2\_190092 and from the European Research Council (ERC) under the European Union's Horizon 2020 research and innovation programme grant agreement No 864361 (CosmicWeb).

\label{lastpage}

\end{document}